\documentclass[10pt]{article} 
\usepackage{iclr2023_conference,times}
\iclrfinalcopy 

\usepackage{amsmath,amsfonts,bm}

\def\eqref#1{equation~\ref{#1}}

\def\1{\bm{1}}

\DeclareMathAlphabet{\mathsfit}{\encodingdefault}{\sfdefault}{m}{sl}
\SetMathAlphabet{\mathsfit}{bold}{\encodingdefault}{\sfdefault}{bx}{n}

\usepackage{amsmath}
\usepackage{amsfonts}
\usepackage{amssymb}
\usepackage[protrusion=true,tracking=true,kerning=true,expansion,final]{microtype} 

\usepackage{multicol}
\usepackage{booktabs}
\usepackage{graphicx}
\usepackage{multirow}
\usepackage{pifont}%
\usepackage{listings}
\usepackage{tabularx}

\usepackage{inconsolata}

\usepackage{paralist}

\usepackage{subcaption}
\usepackage{xparse}

\vbadness=\maxdimen
\usepackage{float}
\newfloat{lstfloat}{htbp}{lop}
\floatname{lstfloat}{Listing}

\usepackage[pagebackref=true,breaklinks=true,colorlinks,bookmarks=true]{hyperref}

\definecolor{citecolor}{rgb}{.259,.659,1}
\definecolor{mydarkblue}{rgb}{0,0.08,0.45}
\definecolor{urlcolor}{rgb}{0,.145,.698}
\definecolor{linkcolor}{rgb}{0.01,0.31,.65}
\hypersetup{
    colorlinks=true,
    breaklinks=true,  %
    bookmarksnumbered=true,
    linkcolor=linkcolor,
    citecolor=citecolor,
    filecolor=mydarkblue,
    urlcolor=urlcolor,
    pdftitle={BigCode Governance Card},
    pdfpagemode=FullScreen,
    pdfview=FitH
    }

\renewcommand*{\backref}[1]{} %
\renewcommand*{\backrefalt}[4]{%
	\ifcase #1 %
	\or
	(cited on p. #2)%
	\else
	(cited on pp. #2)%
	\fi
}
\makeatletter
\renewcommand{\@biblabel}[1]{#1.}
\makeatother

\title{The BigCode Project Governance Card}  

\begin{document}
\maketitle
\begin{center}
\textbf{Sean Hughes}$^{1, \star}$\quad
\textbf{Harm de Vries}$^2$\quad
\textbf{Jennifer Robinson}$^1$\quad
\\\textbf{Carlos Muñoz Ferrandis}$^{3, \star}$\quad 
\textbf{Loubna Ben Allal}$^3$\quad
\textbf{Leandro von Werra}$^3$\quad
\\\textbf{Jennifer Ding}$^4$\quad
\textbf{Sebastien Paquet}$^2$\quad
\textbf{Yacine Jernite}$^3$\quad

$^{1}$ServiceNow \quad
$^{2}$ServiceNow Research\quad
$^{3}$Hugging~Face\quad
$^{4}$The Alan Turing Institute\quad

Corresponding authors ($\star$) can be contacted at \url{contact@bigcode-project.org}
\end{center}

\begin{abstract}

This document serves as an overview of the different mechanisms and areas of governance in the BigCode project.
It aims to support transparency by providing relevant information about choices that were made during the project to the broader public,
and to serve as an example of intentional governance of an open research project that future endeavors can leverage to shape their own approach.       
The first section, Project Structure, covers the project organization, its stated goals and values, its internal decision processes, and its funding and resources.
The second section, Data and Model Governance, covers decisions relating to the questions of data subject consent, privacy, and model release.

\end{abstract}

\section{Project Structure}

\subsection{Goals and Values}

\subsubsection{Project Overview}

BigCode is an open scientific collaboration working on the responsible development and use of large language models for code, aiming to empower the machine learning and open source communities through open governance.

Code LLMs enable the completion and synthesis of code, both from other code snippets and natural language descriptions, and can be used across a wide range of domains, tasks, and programming languages. These models can, for example, assist professional and citizen developers with building new applications.

One of the challenges typically faced by researchers working on code LLMs is the lack of transparency around the development of these systems. While a handful of papers on code LLMs have been published, they do not always give full insight into the development process, which hinders both external accountability and the ability of all but a few well funded research labs to meaningfully participate in shaping the technology.

BigCode is a community project jointly led by Hugging Face and ServiceNow. Both organizations committed research, engineering, ethics, governance, and legal resources to ensure that the collaboration runs smoothly and makes progress towards the stated goals. ServiceNow Research and Hugging Face have made their respective compute clusters available for large-scale training of the BigCode models, and Hugging Face hosts the datasets, models, and related applications from the community to make it easy for everyone to access and use.

An open-invitation was extended to the global AI research community to join forces on the development of state-of-the-art code LLMs, with a focus on research topics such as:
\begin{itemize}
    \item Constructing a representative evaluation suite for code LLMs, covering a diverse set of tasks and programming languages
    \item Developing new methods for faster training and inference of LLMs
    \item The legal, ethics, and governance aspects of code LLMs
\end{itemize}

The BigCode project is conducted in the spirit of open science. Datasets, models, and experiments are developed through collaboration and released with permissive licenses back to the community. All technical governance takes place within working groups and task forces across the community.

As code LLMs are developed with data from the open-source community, we believe open governance can help to ensure that these models are benefiting the larger AI community. We developed tools to give code creators agency over whether their source code is included in the training data, and to find approaches that give attribution to developers when models output near-copies of the training data contained in The Stack.

\subsubsection{Technical and Performance Goals}

The overarching technical goal envisioned before the project was announced was to train and release a 12-billion parameter model that matches Codex \citep{chen2021evaluating}. 
This model from OpenAI was not released and was only available as API service under the name \texttt{code-cushman-001}, although it’s not entirely clear if this model matches the one described in the paper. It has also been suggested by \citep{thakkar2022copilot} that this model is used in Github CoPilot. Our original plan was to compare model performance on HumanEval \citep{chen2021evaluating} and APPS \citep{hendrycks2021measuringapps}, but along the way, we recognized the need for creating an extensive evaluation suite for Code LLMs.

The project ended up breaking the challenge into development phases, starting with the collection of permissively licensed repositories from Github. This initial phase was performed
by the ServiceNow team over several months prior to the official launch of BigCode. It involved inventorying active GitHub repository names, managing the effort to download those repositories, filtering to exclude large files and duplicates, and detecting the licenses used for each repository.  This effort ultimately resulted in the creation of 
The Stack \citep{kocetkov2022stack}, a source code dataset that marked the first milestone for the project.

Two cycles of model development were conducted by the BigCode community. The first cycle took place in November-December 2022, and culminated with the release of SantaCoder, a 1.1B parameter model trained on the Java, JavaScript, and Python code from The Stack. In the next cycle, which was held from January to April 2023, the community scaled up their efforts and trained 15.5B parameter models on 1T tokens from The Stack. The resulting StarCoder models either match or surpass the \texttt{code-cushman-001} model on a variety of coding benchmarks. 

\subsubsection{Social Impact Dimensions and Considerations}

Technical goals and considerations of social impact go hand in hand, and participate equally in \textit{responsible}  development of code LLMs. Within the BigCode project, this means that organizational and technical choices were jointly shaped by the pursuit of the performance goals outlined above and by a best-effort approach to accounting for their impact on various categories of external stakeholders. In particular, participants of the BigCode project focused on the following three dimensions of social impact:

\begin{itemize}
    \item \textbf{Consent of data subjects}: the success of code LLMs depends on their training data, which is the product of the professional and volunteer work of software developers. Since training large models constitutes a novel use and transformation of this work, it poses new questions and challenges with respect to the wishes and rights of the developers.
    \item \textbf{Privacy}: investigations into the behaviors of previous large-scale code LLMs outlined privacy risks when the models can be prompted to generate private information contained in its training data. Addressing these risks was the focus of several related efforts in BigCode.
    \item \textbf{Software safety and security}: recent work by \citep{khlaaf2022hazard} has also shed light on different hazards that are unique to or exacerbated by code LLMs, including their dual use potential in facilitating malware generation or their likelihood of recommending code that includes security flaws.
\end{itemize}

We found that while these considerations did sometimes result in trade-offs between the performance goals and social impact concerns, they were more often better addressed by developing new technical and organizational tools, which we describe in the rest of this document and share as an important outcome of the BigCode project so they can be leveraged by future similar endeavors.

\subsection{Organizers and Participants}
\subsubsection{Inception}

The idea for the BigCode Project came about in Utrecht during a discussion initiated by Harm de Vries (ServiceNow Research) with Thomas Wolf (Hugging Face). Inspired by the BigScience Project, Harm recognized the shared vision of ServiceNow and Hugging Face to responsibly develop open and responsible large language models for code, and approached Thomas to explore the idea of a jointly led open-scientific collaboration with the global machine learning and open source communities. As it turns out, the visions were indeed aligned, and work got started to initiate the project.

A research collaboration agreement between ServiceNow and Hugging Face created the enabling framework for the project, and set out the terms for rallying the broader scientific community at large to work towards developing, training, exploring, and releasing large foundation models for code.

The scope of the collaboration covered the preparation of training data, including developing tools for downloading publicly accessible code data and for running license detectors, developing tools for filtering data sources based on approved licenses and file type, and releasing this training data to the AI community.

The scope also covered the training of dense transformer models that adopt the mechanism of self-attention through the Megatron-LM architecture, training of retrieval augmented code generation models, and developing tools to diagnose instabilities arising from training transformer models at scale.

The collaboration would also prepare an evaluation suite with the help of the scientific community 1) to develop the tools and scripts needed to use existing program synthesis benchmarks such as HumanEval \citep{chen2021evaluating} and CodexGlue \citep{lu2021codexglue}, and 2) to construct and release openly available benchmarks that measure desirable capabilities of large multi-lingual code LLMs and tasks such as program synthesis, text-to-code, and code summarization.

Key milestones identified during the initial stages were focused on developing a community engagement plan, a first attempt at constructing an evaluation suite over multiple programming languages, investigating viable sources of data, and then training and releasing a 12B parameter model that matches Codex performance on HumanEval and APPS.

\subsubsection{Participants}
BigCode is a research collaboration and is open to participants who:
\begin{enumerate}
    \item have a professional research background and
    \item are able to commit time to the project.
\end{enumerate}
In general, we expect applicants to be affiliated with a research organization (either in academia or industry) and work on the technical/ethical/legal aspects of LLMs for coding applications. Throughout the project the community invited guest subject matter experts to participate in certain discussions, and this resulted in a lift in the number of participants in chat channels relative to the number of researchers that had formally applied to participate in the research.

As of May 4th 2023, BigCode has 675 participants with 629 members across the research community (including from Hugging Face and ServiceNow) from 62 countries. The top 5 countries include USA (222), India (60), UK (36), Canada (35), and Germany (30). The community communicates across a total of 48 Slack channels, including Steering Committee (3 channels), Working Groups (7 channels), Task Forces (25 channels), and General Community (13 channels).

Everyone who joins the project is required to follow the BigCode Code of Conduct \citep{BigCodeCodeOfConduct}, understand how we manage intellectual property \citep{BigCodeIP}, and are encouraged to introduce themselves, and to join any working group or task force that aligns to their own interests. If a group does not cover their interests, they are encouraged to pitch their ideas and to take a leadership role for a new working group or task force with the approval of the Steering Committee. 

\subsubsection{Project Governance}
The BigCode project is governed by a steering committee jointly led by Harm de Vries (ServiceNow) and Leandro von Werra (Hugging Face), and supported by a core team comprised of Raymond Li, Denis Kocetkov, and Sean Hughes from ServiceNow, and Carlos Muñoz Ferrandis, Loubna Ben Allal, and Thomas Wolf of Hugging Face. Through the course of the project, additional members were added to the core team, including Yacine Jernite, Armel Randy, and Joel Lamy-Poirier.

The Steering Committee is effectively responsible for organizing and managing the project (including research strategy and publication goals), and provides oversight across all working groups. Decisions that cannot be addressed at the community level would be elevated to the lead of the Working Group for facilitated discussion, with further inputs and tie-breaker decision making by the Steering Committee as a last resort. Governance for the project is open, meaning that the BigCode project encourages anyone from the community to join any working group or task force of interest, and for them to engage and contribute to work and decision making in the group.

\subsubsection{Timeline, Milestones, and Community Events}
The BigCode project was announced on September 26, 2022 \citep{BigCodeProject2022Sep26}. We shared the goal of the project to train a state-of-the-art~15B parameter language model for code that will be trained using the \href{https://www.servicenow.com/research/}{ServiceNow Research} in-house GPU cluster. With an adapted version of Megatron-LM, we planned to train the large model on distributed infrastructure. Throughout the project's progress, updates were regularly published in various media as we describe in what follows.

On October 6, 2022, ServiceNow and Hugging Face held a webinar with the BigCode Community to provide strategic direction for the project and research goals. \citep{BigCodeProject2022Oct6}

On October 27, 2022, we introduced The Stack, a large dataset of more than 3 TB of permissively licensed source code. \citep{BigCodeProject2022Oct27} Our paper \citep{kocetkov2022stack} described the details of the dataset collection, presented a brief dataset analysis, and showed promising results on the HumanEval benchmark. Our experimental results show that near-deduplication is an important pre-processing step for achieving competitive results on text2code benchmarks. We released all permissively licensed files for 30 common programming languages, along with a near-deduplicated version.

On November 15, 2022, we introduced a new tool called “Am I in The Stack” that allows developers to check whether any data from their GitHub repositories is included in The Stack. \citep{BigCodeProject2022Nov15} We also introduced v1 of the BigCode Opt-Out process, which gives agency back to Developers by providing a way for them to request that their data be removed from the dataset.

On November 23, 2022 Loubna Ben Allal shared details of our initial approach for how we planned to tackle the de-identification to remove personally identifiable information (PII) from The Stack. \citep{LoubnaBenAllal2022Nov23}

On November 29, 2022, we shared the Weights and Biases dashboards for our first models so that the broader community could follow along. \citep{BigCodeProject2022Nov29}

On December 1, 2022, we released The Stack v1.1, with 358 programming languages included, and more than double the data, going from 3Tb to 6.4TB, with the help of the legal tech community that identified 193 viable permissive source code license types. Before releasing v1.1 we also removed the first batch of repositories based on opt-out requests. \citep{BigCodeProject2022Dec1}

On December 2, 2022, we held an in-person meetup alongside NeurIPS 2022 in New Orleans with more than 75 members of the BigCode community, where we were able to make the connections to foster greater awareness and understanding of the BigCode project. \citep{BigCodeProject2022Dec2}

On December 9, 2022, a member of the BigCode community held a similar meetup at EMNLP 2022 in Abhu Dhabi, another opportunity to raise awareness of BigCode and to discuss our project with the NLP research community.  \citep{BigCodeProject2022Dec9}

On December 12, 2022, we sent out another message to raise awareness of “Am I in The Stack” and to inform developers about the option to opt-out from the dataset. \citep{BigCodeProject2022Dec12}

On December 14, 2022, Hugging Face and ServiceNow held a second webinar with the BigCode Community to review progress and provide an update on plans for ongoing research towards the 15B parameter model. \citep{BigCodeProject2022Dec14}

On December 22, 2022, we released \href{https://huggingface.co/bigcode/santacoder}{SantaCoder}, a 1.1B multilingual large language model for code that outperforms much larger open-source models on both left-to-right generation, and infilling. \citep{BigCodeProject2022Dec22} The SantaCoder models are licensed under an open and responsible AI model license, CodeML OpenRAIL-M v0.1 \citep{BigCodeOpenRAILLicense}. These are AI-specific licenses enabling free use and distribution of the model while setting specific use restrictions (e.g. malware generation). We published a detailed technical report \citep{allal2023santacoder} that included details of all the key contributions to the development of the model.

On February 1, 2022, members of the BigCode core team were invited to meet with the European Parliament Innovation Lab. \citep{BigCodeProject2022Feb1} At that meeting we shared details \citep{Utopiah2022Feb1} of the project and answered questions from members of the Lab. Engaging with policymakers and regulators is an important part of the journey to inform and educate key stakeholders from the broader AI ecosystem.

On March 20, 2022, we announced The Stack v1.2 which included The Stack Issues, The Stack Metadata, and The Stack Commits. \citep{BigCodeProject2022Mar20} With this release, we simplified the opt-out process and also removed opt-outs from developers where the request was received by February 2023. Along with this release, we provided access to a dataset of GitHub issues totalling 54GB, and we applied the same opt-out mechanism to these issues. The GitHub issues dataset is more conversational and could be helpful to train models to be used as coding assistants.

On April 13, 2023, inspired by discussions in the training working group,  Harm de Vries shared an analysis of Chinchilla scaling laws on how much additional compute resources are needed to create smaller LLMs. \citep{HarmDeVries2023Apr13} 
 These insights suggest we have not reached the limit of training smaller models on more tokens - an important consideration for future research.

On May 4, 2023, BigCode announced StarCoder and StarCoderBase, two code LLMs trained on permissively licensed data from GitHub, including from 80+ programming languages, git commits, GitHub issues, and Jupyter notebooks. \citep{BigCodeProject2023May4} Similar to \href{https://ai.facebook.com/blog/large-language-model-llama-meta-ai/}{LLaMA} \citep{touvron2023llama}, StarCoderBase is a ~15B parameter model trained on 1 trillion tokens. On top of StarCoderBase a variant called StarCoder is trained for 35B additional tokens purely on Python.

\subsubsection{Supporting Resources and Funding}
Understanding the costs of a project like BigCode can help ground conversations about the trade-offs involved in the development of code LLM technology more broadly, helping understand how various private and public institutions may participate in this development and allocate resources to maximize its overall benefits. We outline the major costs in terms of computation resources, human participation, and organization.

\textbf{Data collection} ServiceNow handled the data collection effort to constitute a raw dataset containing 5.28B files with a total size of 92 TB and filtered it down to build The Stack.

\textbf{Compute and emissions} We trained SantaCoder on the ServiceNow cluster using 96 Tesla V100 GPUs, and StarCoder on a Hugging Face GPU cluster with 512 A100 80GB GPUs distributed across 64 nodes.

We report the carbon footprint of training these models:

\begin{itemize}
    \item SantaCoder: Based on the total number of GPU hours that training took (14,284) and an average power usage of 300W per GPU, this adds up to 4285 kWh of electricity consumed during the training process. Multiplied by the carbon intensity of the energy of the Montreal location  (0.029 kgCO2e per kWh) and assuming an average Power Usage Effectiveness of 1.2,  this results in 124 kg of CO2eq emitted. 
    \item StarCoderBase:  320,256 GPU hours; 280W per GPU; 89671.68 kWh of electricity. Carbon intensity of the energy of the us-west-2 AWS location: 0.15495 kgCO2e per kWh; average Power Usage Effectiveness across AWS datacenters: 1.2. Total emissions: 16.68 tonnes of CO2eq.
\end{itemize}

\textbf{ServiceNow and Hugging Face employees working on BigCode} The estimated time commitment for the duration of the project for employees of the host institutions corresponds to 6 full-time employees for the duration of the project.

\textbf{Estimated volunteer hours across the project} The time commitment from volunteers is harder to estimate given the large number of participants and the variety of time investments across phases and participants. At a minimum, we estimate overall time commitment from volunteers matched time commitment from employees of the host institutions.

\textbf{Community events and appreciation} ServiceNow and Hugging Face organized a community meetup that coincided with NeurIPS 2022 in New Orleans, USA. The budget for the event was approximately \$6,000 from ServiceNow Research for the venue with hospitality. Hugging face also provided promotional items including stickers and tshirts at the event, and sent named contributors to the research paper complimentary BigCode branded tshirts.

\textbf{Data annotation} Hugging Face funded the data annotation services from Toloka, with a total outlay of \$39,000 paid to crowd workers. Since this was a research project, Toloka provided free consulting and agreed to waive the fees for running the annotation tasks on their platform.
\section{Data and Model Governance}

\subsection{Data Governance}

\subsubsection{Data Collection and Management Plan}

In the course of the BigCode project, we collected two main datasets. The primary training dataset is The Stack, which was obtained by gathering public code files, issues, and commits from GitHub. To collect Github repositories, we first extracted a list of repositories from GHArchive \citep{GHArchive} and subsequently cloned all of them using a large CPU cluster. We also used the data from GHArchive to extract the Github issues. The git commits were gathered from a public BigQuery service. Additionally, we collected a dataset of annotations of several kinds of private information on a subset of The Stack to support our privacy risk mitigation efforts.

The legal basis for data collection under fair use and with regards to GDPR and the corresponding case law are still evolving. In this context, the data collection and data management plans were carefully crafted with support from leading experts in the open source and legal tech community that participated in the Legal, Ethics, Governance Working Group in a best-effort approach to reflect current understandings of legal requirements for data collection and management.

\textbf{The Stack Dataset Access and Management} The StarCoder model was trained on The Stack v1.2, which exclusively contains 6.4TB of permissively licensed data \citep{BlueOak} from GitHub repositories, processed from an original source dataset of 102TB. Access and management follow the following schema:

\begin{itemize}
    \item \textbf{What data can be accessed:} the 6.4TB of processed data can be accessed through the Hugging Face Hub, while the original 102TB are only accessible to the stewards of the project  for the purposes of enabling the research and to support future internal and external requirements that may arise, for example to search the full dataset to recall licenses, determine code provenance, and attribution.
    \item \textbf{What are the conditions for accessing the data:} users are able to inspect the dataset via the Dataset Card and embedded Dataset Preview, but are required to agree to the Terms of Use for The Stack \citep{TheStackTOU} before being able to download it. This includes the requirements to 1) abide by the terms of original source code licenses, including attribution clauses when required (The Stack provides provenance information for each data point), 2) agree to update copies of The Stack to the most recent usable version specified \href{https://huggingface.co/datasets/bigcode/the-stack/discussions/7}{here}, and 3) include the Terms of Use and require users to agree to it if a copy is to be hosted, shared, or otherwise provided. As of May 3, 2023, The Stack had been downloaded 50,200 times. 
    \item \textbf{How can a data subject request that their data be removed:} we provide an opt-out form that lets people opt out of having any code or text they put on GitHub be included in The Stack. Additionally, anyone who is concerned about specific data they have encountered in The Stack, for example relating to PII, malicious code, or code that has an incorrect license or attribution can email \texttt{contact@bigcode-project.org}. At the time of the data processing for the StarCoder model training, 44 people had opted out of The Stack and associated repositories were removed.
    \item \textbf{How often is the data updated:} For as long as we are maintaining The Stack dataset, we will provide regular updates to the dataset to remove data that has been flagged since the last version. This includes data that has been opted out, and data that was flagged as containing PII, malicious code or using a non-permissive license since the previous release. The current plan is to update the dataset every 3 months, although the schedule may change based on the volume of requests received. If we are not in a position to continue maintaining the dataset, we plan to stop distributing it in its current format and update its terms of use to limit its range of applications further.
\end{itemize}

\textbf{PII Dataset Access and Management} In order to support our efforts to mitigate the risk that the model may leak private information, we selected 12,000 samples of code from The Stack and annotated them to detect PII using crowd-sourcing. The resulting dataset  was used to train a PII detection model that we used to detect and then mask PII (Names, Emails, IP addresses, Keys, Passwords) from our StarCoder training dataset.

\begin{itemize}
    \item \textbf{What data can be accessed:} the data is hosted as a gated dataset on the Hugging Face Hub. The dataset will be made available to researchers on a case-by-case basis for research projects that require access, in addition to the original team who developed the dataset.
    \item \textbf{What are the conditions for accessing the data:} researchers who want to access the dataset need to request access and be approved by the maintainers as well as agree with the dataset's Terms of Use
    \item \textbf{How can a data subject request that their data be removed:} as a derived dataset of The Stack, the PII dataset will be updated to reflect data that has been opted out from the source dataset.
    \item \textbf{How often is the data updated:} similarly, following The Stack terms of use, the PII Dataset will be updated as often as the Stack if some of the files it contains have been opted out.
\end{itemize}

\subsubsection{Consent of Data Subjects}
\textbf{Between implicit and explicit consent} One of the goals of BigCode is to give developers agency over their source code and let them decide whether or not it can be used to develop and evaluate LLMs. Software developers typically rely on licenses to express how they want their work to be re-used; in particular, developers who choose Open Source licenses often do so because they want their code to be broadly re-used. This motivated us to start by selecting data from repositories that met the following criteria:

\begin{itemize}
    \item The repository has an open source license attached - open source, while chosen for very different reasons by different people, typically indicates a willingness to have one's work reused or adapted
    \item The license does not have an attribution clause - attribution is a difficult technical problem for code LLMs. Since we cannot guarantee that the model will be used in a way that attributes its generations to specific training data in a way that satisfies the intent of the licensor, we chose to only keep licenses without an attribution clause
\end{itemize}

Selecting repositories based on licenses is only the first step, however, as many of these licenses were chosen before the recent developments in code LLMs. Thus, we complement this initial approach by also giving repository owners the ability to \textbf{opt out} of having their repositories included in The Stack. We see this approach as a meaningful step forward in improving the agency of data subject in the development of code LLMs, and we present both the tools we developed to support it and its known limitations in the rest of this section.

\textbf{Technical tools to support opt-out} We developed a tool called Am I in The Stack \citep{InTheStack} to help developers inspect The Stack dataset and for them to see whether any of their repositories have been included and that might be used for training LLMs. If that is the case, we show them a custom link that allows them to easily send a request through GitHub, in two clicks if they are already logged in. We chose to mirror the original platform governance by letting the repository owner decide whether code in a repository is included or not in the dataset. This also allows us to validate requests automatically, since the GitHub username \citep{bigcode_data_removal} must match the one used to submit the request. Validated requests and associated code pointers are stored so that the code does not appear in future versions of The Stack.

\textbf{Community feedback on the approach} In the period January-March 2023, members of the BigCode project conducted community research with individuals at specific organizations whose data is used in The Stack, namely The Alan Turing Institute and The Turing Way \citep{turing_way_2022} as well as two open, international workshops Open Data Day 2023 \citep{open_data_day_event} and \href{https://schedule.mozillafestival.org/session/KAS9YF-1}{Mozilla Festival 2023} with a session titled ‘Designing for Data Rights in the AI Production Pipeline’. These qualitative interviews and participatory co-design workshops included 50 participants primarily from North America and Europe with roles like research scientist, community manager, software engineer, and principal investigator (PI).

The outcomes from the community research can be summarized as follows: when it comes to governance of LLM datasets, participants feel that it is both \textbf{better to know} AND \textbf{better to have a choice}. Most participants had neutral to positive feelings about their permissively licensed data being used to train LLMs. While all had positive impressions of the ``Am I in The Stack'' tool, no one interviewed expressed a desire to actually opt-out. The main takeaway seemed to be that participants found the most value in BigCode governance tools for their ability to raise awareness of data practices and to empower individuals and communities to take actions based on their specific needs. The co-created outputs can be viewed on the MozFest Miro Board \citep{open_data_day_miro}.

Additionally, during the first stage of the opt-out process, individuals \textbf{who chose to have their data removed from the Stack} were asked to specify the reasons for wanting their code to be excluded from the dataset. The responses revealed a few recurring themes, including:

\begin{itemize}
    \item Preference for an opt-in approach instead of opt-out
    \item Perception that it is unfair to use their code without compensation
    \item Concerns about the current limitations of AI and the potential for model generations to be traced back to their work, resulting in potential legal liability.
    \item Belief that their code is of poor quality and unsuitable for AI training.
    \item Presence of PII in their code, which they do not wish to be publicly exposed.
\end{itemize}

The feedback form also revealed another limitation of the opt-out process. When code is licensed permissively or under a copy-left license, it can be duplicated to another repository, making it challenging to eliminate such copies if the copyright owner chooses to opt-out. More work is necessary to create workable data control and consent mechanisms for the large-scale training data of LLMs. 

\subsubsection{Private Information Handling}
One significant concern with respect to privacy was the risk that the code LLM may generate private information found in its training data, including private tokens or passwords matched with identifiers or email addresses. Additionally, while users can (and have) requested that data be removed from The Stack dataset because it contains personal data, removing specific information from trained model weights after the fact remains an open technical challenge. In order to minimize this risk, we chose to apply automated PII redaction at the pre-processing stage during training.

Our first step toward automatic PII redaction consisted in creating an annotated dataset for PII in code data, as we found that neither regular expression-based approaches nor existing commercial software for PII detection met our performance requirements. In doing so, we aimed to balance the constraints of costs (fair compensation), time (the timing and time to complete the work was on the critical path for the project), and quality (to ensure that PII Detection Model training was not impacted). While traditional data annotation services using salaried employees were considered, we decided to work with crowd-workers through Toloka after reviewing several service providers and their compensation practices - and finding that most would not provide sufficient transparency and guarantees about worker compensation. We selected pay and eligible countries of crowd-workers to ensure that 1. the absolute hourly wage was always higher than the US federal minimum wage (\$7.30), and 2. the hourly wage was equivalent to the highest state minimum wage in the US in terms of purchasing power parity (\$16.50 at the time of writing).

We engaged 1,399 crowd-workers across 35 countries in annotating a diverse dataset for PII in source code. Our PII detection model, trained on 22,950 secrets, achieves 90\% F1 score surpassing regex-based tools, especially for secret keys. The PII annotations are available to approved individuals, and researchers and developers that are granted access are expected to uphold ethical standards and data protection measures. By making it accessible, our aim is to encourage further research and development of PII redaction technology.

We also released \textbf{StarCoderData}, the pre-processed version of The Stack used to train the StarCoder model, which has its PII redacted using our model.

\subsection{Model Governance}

\subsubsection{Model Licensing}
The model is released under an open and responsible AI model license agreement (BigCode OpenRAIL-M v1.0 \citep{BigCode_OpenRAIL_License_v1} which enables royalty free access and flexible use and sharing of it, while setting specific use restrictions for identified critical scenarios.  Most importantly, the license agreement requires stakeholders wishing to share the model or a modified version of it: (i) to include the same set of use restrictions or a similar one in their legal agreements; (ii) to keep the model card and provide a similar one or one of better quality when sharing a modified version of the model (FAQ for the model license agreement \citep{BigCode_OpenRAIL_FAQ}).

The BigCode OpenRAIL-M license agreement (i.e. the legal document itself) is available under a CC-BY-4.0 license. Therefore, any stakeholders can freely adopt the same license agreement for their models, or modify it for their specific AI artifacts. For more information about responsible AI licensing, please visit the RAIL Initiative webpage, The Turing Way Handbook for ML researchers \citep{turing_way_licensing_ml}, or OECD AI content on RAILs and trustworthy AI principles \citep{oecd_ai_trustworthy}.

\subsubsection{Attribution Tool}
With SantaCoder we released a tool for developers to check whether generated source code had been trained on data from The Stack, and if so, the tool would return the likely matches with full attribution. We both offer a fast membership test to check if code was part the pretraining data as well as a full-text search tool. With StarCoder we are releasing a similar tool (StarCoder Dataset Search \citep{huggingface_bigcode_search}) enabling users to check the origin of the model output and respect any licensing conditions (if any).

\section{Conclusion and Acknowledgements}
\subsection{This is a snapshot}
Please note that this is a snapshot of the evolving governance of the BigCode project. A more up-to-date view may be found in the \href{https://huggingface.co/datasets/bigcode/governance-card/}{BigCode Governance Card} on the Hugging Face website. The intention is to add more details about the project over time. Please leave us comments in the \href{https://huggingface.co/datasets/bigcode/governance-card/discussions}{Community} if there are any questions or requests for more insights about the project governance. Thank you for taking the time to read this document. We hope it is useful.

\subsection{Acknowledgements}
The work presented in this card is the outcome of the efforts of many BigCode participants beyond the authors of the card.
Please refer to the published papers detailing this work for contributions,
e.g. StarCoder \citep{li2023starcoder},
The Stack \citep{kocetkov2022stack}, and SantaCoder \citep{allal2023santacoder}.

\clearpage

\bibliography{bigcode}
\bibliographystyle{iclr2023_conference}

\end{document}